\documentclass[letterpaper]{article} 
\usepackage{aaai2026}  
\usepackage{times}  
\usepackage{helvet}  
\usepackage{courier}  
\usepackage[hyphens]{url}  
\usepackage{graphicx} 
\usepackage{natbib}  
\usepackage{caption} 
\usepackage{booktabs} 
\usepackage{amsmath}
\usepackage{amsfonts}
\usepackage{multirow}
\usepackage[flushleft]{threeparttable}
\usepackage{makecell}
\usepackage{subcaption}
\usepackage{algorithm}
\usepackage{algorithmic}

\usepackage{newfloat}
\usepackage{listings}
\DeclareCaptionStyle{ruled}{labelfont=normalfont,labelsep=colon,strut=off} 
\floatstyle{ruled}
\newfloat{listing}{tb}{lst}{}
\floatname{listing}{Listing}
\title{Save, Revisit, Retain: A Scalable Framework for Enhancing User Retention in Large-Scale Recommender Systems}
\author{
    Weijie Jiang,
    Armando Ordorica,
    Jaewon Yang, \\
    Olafur Gudmundsson,
    Yucheng Tu\textsuperscript{\rm}\thanks{Work done at Pinterest.},
    Huizhong Duan
}
\affiliations{

    Pinterest Inc.\\
    {weijiejiang,aordorica,jaewonyang,ogudmundsson,ytu,hduan}@pinterest.com
}

\usepackage{bibentry}

\begin{document}

\maketitle

\begin{abstract}
User retention is a critical objective for online platforms like Pinterest, as it strengthens user loyalty and drives growth through repeated engagement. A key indicator of retention is revisitation, i.e., when users return to view previously saved content, a behavior often sparked by personalized recommendations and user satisfaction. However, modeling and optimizing revisitation poses significant challenges. One core difficulty is accurate attribution: it is often unclear which specific user actions or content exposures trigger a revisit, since many confounding factors (e.g., content quality, user interface, notifications, or even changing user intent) can influence return behavior. Additionally, the scale and timing of revisitations introduce further complexity; users may revisit content days or even weeks after their initial interaction, requiring the system to maintain and associate extensive historical records across millions of users and sessions. These complexities render existing methods insufficient for robustly capturing and optimizing long-term revisitation.

To address these gaps, we introduce a novel, lightweight, and interpretable framework for modeling revisitation behavior and optimizing long-term user retention in Pinterest’s search-based recommendation context. By defining a surrogate attribution process that links saves to subsequent revisitations, we reduce noise in the causal relationship between user actions and return visits. Our scalable event aggregation pipeline enables large-scale analysis of user revisitation patterns and enhances the ranking system’s ability to surface items with high retention value. Deployed on Pinterest’s Related Pins surface to serve 500+ million users,
the framework led to a significant lift of 0.1\% in active users without additional computational costs. 
Our data analysis reveals novel insights, such as the impact of content topics on revisitation rates; for example, users are more likely to revisit aesthetically pleasing topics.  
\end{abstract}


\section{Introduction}


User retention is an important business objective for online platforms such as Pinterest, as it fosters stronger relationships with users, builds loyalty, and ultimately drives both growth and revenue. A key behavior indicating strong user retention is revisitation \cite{zhang2021user}, which occurs when a user returns to view specific content that they previously accessed and saved. This behavior often signals user satisfaction, typically driven by personalized content and recommendations. Therefore, identifying and leveraging previously saved content that encourages users to return can serve as a valuable signal for fostering long-term user retention. 

Modeling and optimizing revisitation behavior presents several challenges. The first is attribution: it is difficult to determine what exactly triggers a user to revisit. Many confounding factors can influence revisitation, such as content quality, user experience, and platform notifications or reminders \cite{adar2008large, zhang2021user}. For instance, if a user returns with a completely different intent than in their previous session, it would be inappropriate to attribute that revisit to their earlier activities.
The second challenge concerns scale. Since revisitations can occur within a day, a week, or even a month~\cite{adar2009resonance}, tracking them effectively requires maintaining extensive records of user activity over indefinite periods. This creates significant data processing and scaling challenges~\cite{jin2017personal}. To date, there has been little concrete research on whether it is reasonable to disregard revisitations after a certain period (e.g., after X days).



Because of these challenges, many existing methods prioritize immediate engagement metrics (e.g., clicks, saves) over longer-term retention \cite{gugnani2020implicit, wang2020discovery, dang2025augmenting, gong2025multiple, jia2025learn}. In some cases, effective modeling also increases the time users spend \cite{guo2022intelligent}. Reinforcement learning (RL) offers a framework for optimizing long-term rewards. Although RL methods have improved intra-session engagement \cite{afsar2022reinforcement, cai2023reinforcing, liu2024modeling}, applying RL to optimize longer-term outcomes such as revisitation (events that may occur days after exposure) remains highly challenging. Existing RL approaches often lack the scalability and robustness to handle multi-day sequences that may encompass thousands of recommended items. To the best of our knowledge, there are currently no RL-based approaches that, at Pinterest’s scale, explicitly target long-term revisitation while jointly addressing attribution and scalability.

\begin{figure}[t]
  \centering
  \includegraphics[width=\linewidth]{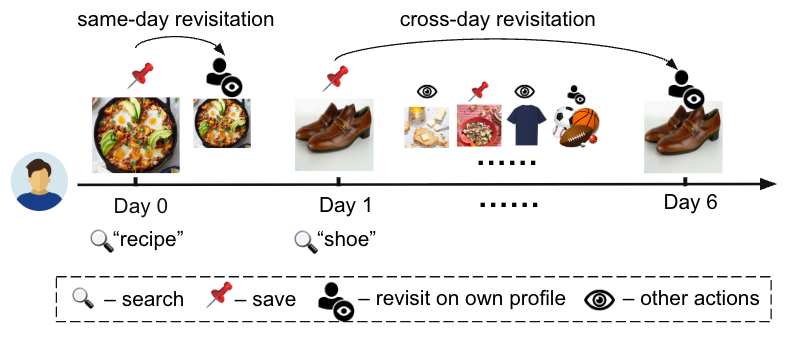}
  \caption{Revisitation Attribution Illustration}
    \label{revisitation_intro}
\end{figure}

In this study, we propose a novel lightweight and interpretable framework for modeling user revisitation behavior and optimizing for long-term user retention in the recommendation context of search. Research shows that well‑chosen user behaviors can serve as effective surrogate rewards for predicting and optimizing long‑term engagement in recommender systems \cite{wang2022surrogate}. 
Therefore, to address the first challenge of attribution, we establish a surrogate for the causal effect of saving an item on subsequent revisitation events to reduce ambiguity and noise in revisitation signals. Figure \ref{revisitation_intro} illustrated two types of revisitation attribution process.

\begin{itemize}
    \item \textit{Same day revisitation: A user searched for ``recipe" on day 0 and saved a particular recipe item to their profile. Later that day, they revisited the saved recipe item in their profile. In this case, we attribute revisitation to the saved recipe item. We do not attribute other items the user viewed on other days to the saved recipe item.}
    \item \textit{Cross-day revisitation: The user saved a shoe item via search on day 1 and viewed some other items in subsequent days. We do not attribute revisitation to the shoe item until the user revisited it in own profile on day 6.}
\end{itemize}
To leverage the constructed revisitation attributions and address the second challenge of scale, we first conducts a comprehensive analysis of user revisitation patterns and then translate the findings into a scalable data pipeline. The pipeline aggregates event data over a seven‑day window across sessions and presentation surfaces by linking search‑surface save events to profile‑surface revisit events. This aggregation allows us to capture the most important revisitation signals for saved items, which serve as an auxiliary prediction task for the ranking model.
In this way, the recommender system prioritizes items with high predicted probabilities of saving and revisit in the top positions, thereby driving downstream return visits and promoting long-term retention. 

This framework was shown to be effective in significantly increasing active users by 0.1\% through both extensive offline experiments and a large-scale online A/B test with 24 million users on Pinterest's Related Pins surface \cite{liu2017related} over two months. 
In summary, our contributions include:
\begin{itemize}
    \item \textbf{Large-scale analysis}: We provide the first large-scale analysis of user revisitation patterns on online platform with hundreds of millions of users.
    \item \textbf{Methodology \& metrics improvements}: We propose a novel, lightweight, and scalable framework for modeling user revisitation behavior without additional cost, which is demonstrated to improve long-term user retention via a large-scale online experiment of 24 millions of users. 
    \item \textbf{Intepretability \& insights}: The proposed framework offers greater interpretability than the SOTA methods, and it is able to show the reasoning behind the metrics improvement and the long-term and short-term revisitation patterns on topics. 
    \item \textbf{Deployment}: The proposed method has been deployed on Pinterest’s Related Pins surface \cite{liu2017related}, efficiently serving 500+ millions of users without incurring additional computational costs.. 
    
\end{itemize}

\section{Related Work}
User retention is critical objective for online platforms as sustained engagement typically drives more growth and revenue than short-term interactions. However, typical recommender systems in online platforms mainly focused on immediate engagement metrics such as click-through rates but struggled to capture the
temporal dynamics essential for user retention. The first challenge is that there are lots of noise in the collected data of user behaviors when modeling retention and it is difficult to find the factors that contribute to user retention \cite{sun2025they}. \citet{ding2023interpretable} designed a contrastive multi-instance learning framework to explore the rationale and improve the interpretability of user retention, and they argue that recommender systems should rank different items for a user according to the user-item retention scores. The second challenge is the supervised information of user retention has been expected to be much sparser than immediate explicit feedback, such as click and save. Generative Flow Networks \cite{liu2024modeling} have been proposed to model retention by treating it as a probabilistic flow over user sessions, which aims to address the challenges of sparse and delayed retention signals. DT4Rec \cite{zhao2023user} leveraged the Decision Transformer to optimize long-term user retention by modeling sequences of interactions. This approach incorporates an autodiscretized reward prompt that preserves partial order relationships between rewards, enhancing retention modeling over time. 

Other workstream such as customer purchase
behavior analysis in industries like fashion have empirically validated that diverse product recommendations not only increase purchase rates but also positively influence long-term retention \cite{kwon2020art}. Fashion is one of the most popular topic on Pinterest. However, our past experiments when intentionally improving recommendation diversity hurt engagement metrics like click and save by more than $1\%$ and we observed the tradeoff tension between relevance and diversity. 

\citet{wang2022surrogate} demonstrated that well‑chosen user behaviors can serve as effective surrogate rewards for predicting and optimizing long‑term engagement in recommender systems. Through deep analyses on the revisitation behaviors of users on Pinterest, we observed that user revisitation behaviors are less sparse than some other user behaviors such as long click and share. Therefore, we establish our surrogate as the causal relationship between a saved item and its associated revisitation events. Although \citet{zhang2021user} designed a separate model for modeling the probability of click $\rightarrow$ revisit and used it together with the ranking model that predicts click rate for predicting retention, deploying an additional model is very costly and increases serving latency. 
In this work, we propose a revisitation modeling framework that jointly models users’ save and revisitation actions within the existing multi‑task recommender, adding no cost or latency relative to the current system.

\section{Revisitation Behavior Analysis}
Users engage in a variety of behaviors on online platforms, such as browsing, clicking on grids, liking, saving, sharing, and commenting. They can also return to the platform for many reasons, from upper-funnel activities like aesthetically-pleasing content to completing lower-funnel projects like DIY and Crafts, or performing outbound clicks on previously saved outfits. Among these behaviors, revisitation stands out as a special type of return behavior, where a user goes back to content they had previously saved, an action that arguably indicates user satisfaction.

In this work, we leverage the common “save” action on online platforms to build an attribution between saved content and subsequent revisitation behaviors. Specifically, when a user saves an item, it is added to their personal profile (such as a collection or board). Users can then revisit their saved items in two primary ways:
\begin{itemize}
\item {\textbf{Impression-Based Revisitation}}: Scroll or navigate through the saved content without deeper interaction on their personal profile (such as collection or board). 

\item {\textbf{Grid-click-based Revisitation}}: Tap on the saved content to access more information on their personal profile (such as collection or board). 
\begin{figure}[t]
  \centering
  \includegraphics[width=1.0\linewidth]{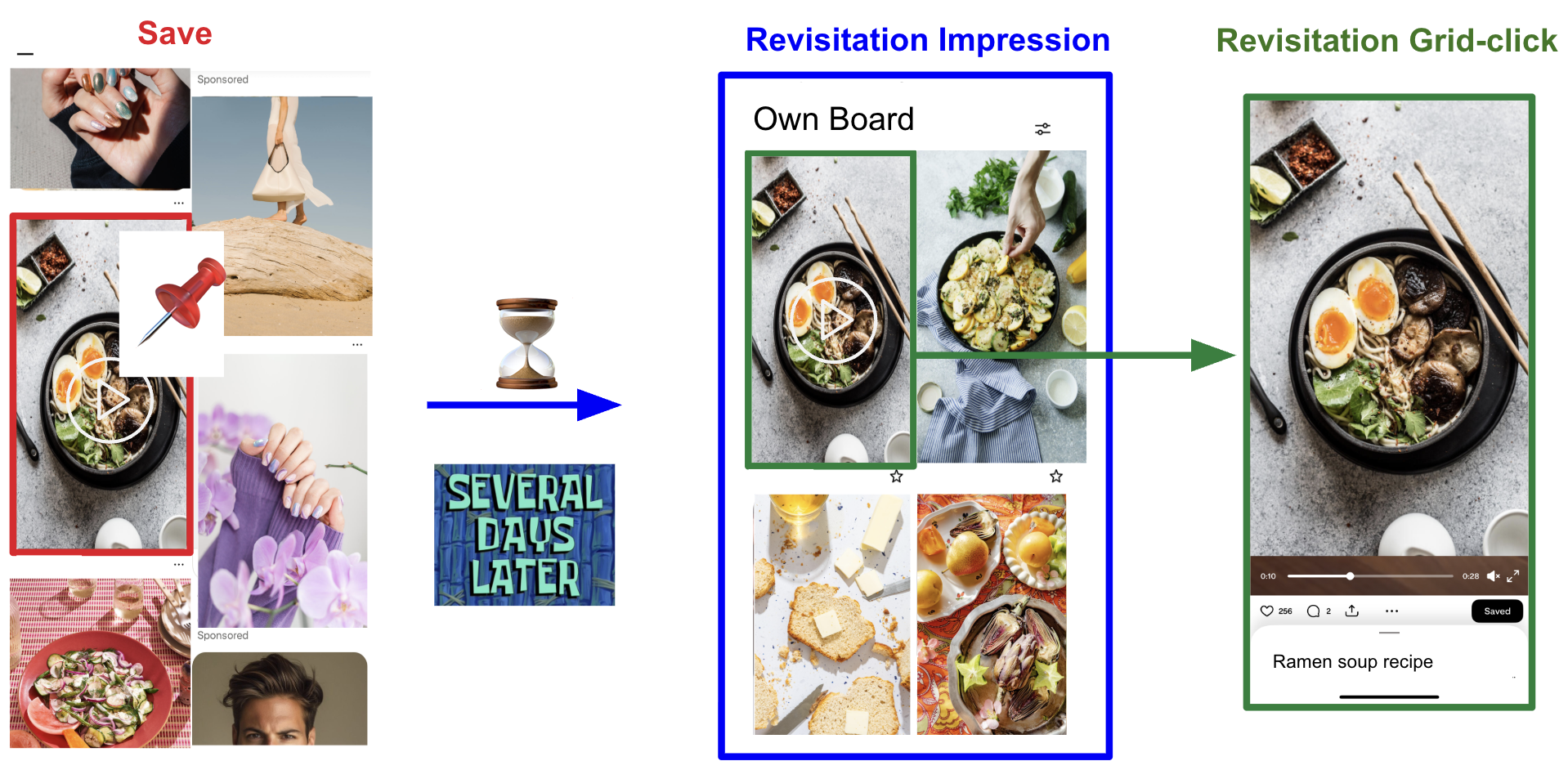}
  \caption{Illustration of Save, Revistation Impression and Revisitation Grid-click on Pinterest}
    \label{re_cu}
\end{figure}
\end{itemize}

Figure \ref{re_cu} shows an example of a user who saved a Pin of a ramen recipe on Pinterest, had a revisitation impression on the Pin on their own board several days later, and then a revisitation grid-click to zoom in the Pin for more details.

We conducted extensive analyses on these two revistation behaviors and observed the following patterns:
\begin{enumerate}
    \item \textbf{More users tend to have revisitation impressions rather than revisitation grid-clicks, and the percentage of users revisiting decays (Figure \ref{a})}. Suppose a user saved a Pin on day 0, 14.6\% of the users tend to have a revisitation impression on the saved Pin on day 0 immediately and 19.5\% of the users on day 1, following a significant decay till 8.7\% on day 9. The percentage of users have a revisitation grid-click starts at 6.6\% at day 0 and peaked at day 1 of 7.7\%, and decays to only 1.6\% day 9.

\item \textbf{The volume of revisitation grid-clicks decays dramatically in the first three days (Figure \ref{b}).} Only 4.7\% of the saved Pins get a revisitation grid-click on day 0 and the percentage dropped by half on day 1 followed by another half on day 3. The percentage dropped more steadily until it reaches 0.5\% by day 5 and 0.3\% by day 9.

\begin{figure*} [t]
\centering

\begin{subfigure}{.45\textwidth}
  \centering
  \includegraphics[width=0.9\linewidth]{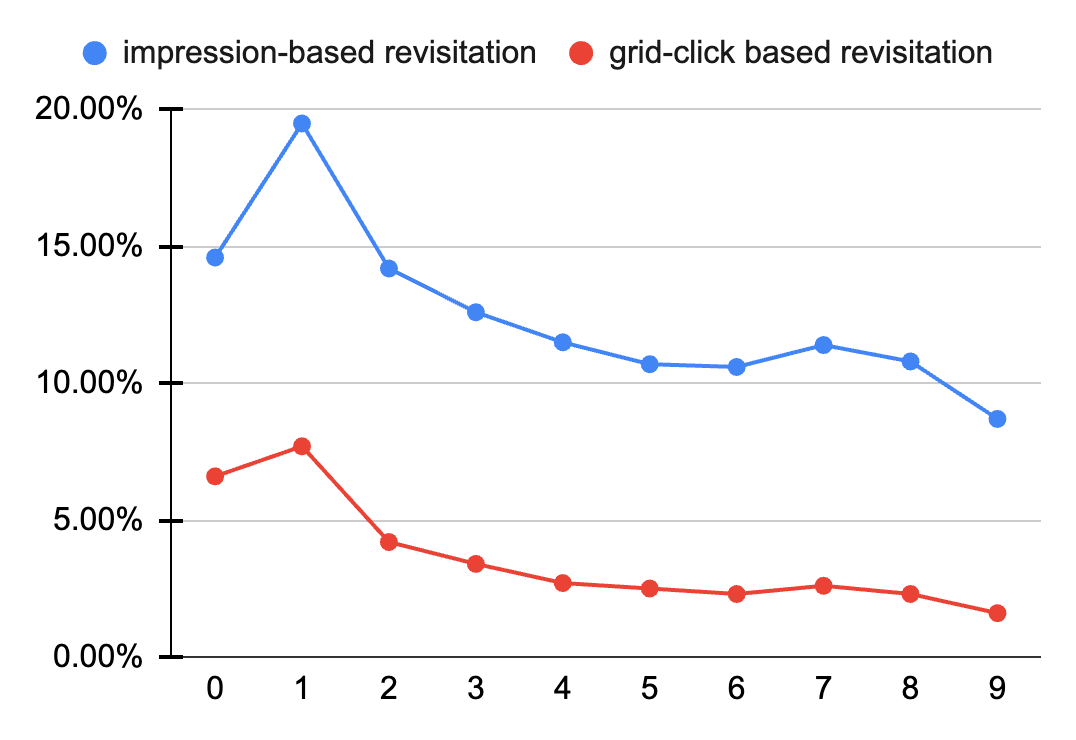}
  \caption{Percentage of Users Revisiting Per Day }
  \label{a}
\end{subfigure}%
\begin{subfigure}{.45\textwidth}
  \centering
  \includegraphics[width=0.9\linewidth]{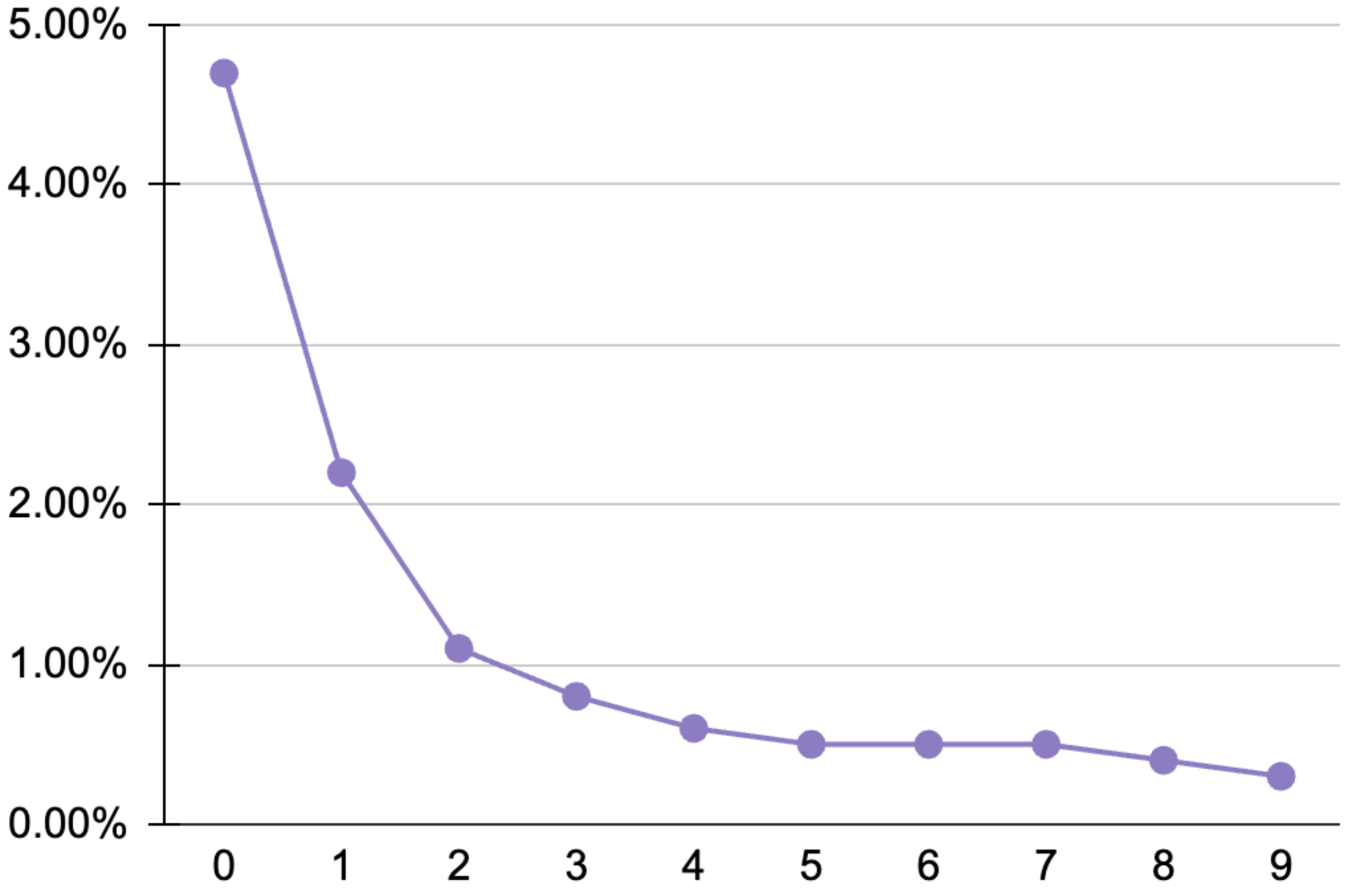}
  \caption{Percentage of Grid-click based Revisitation Volume Per Day}
  \label{b}
\end{subfigure}
\label{}
\caption{Revisitation analysis (users \& volume) - Day 0 is when a Pin was saved}
\end{figure*}


\item \textbf{The sooner they revisit, the higher the expected number of days active in the next 1-month period.} 
Figure \ref{28d-revisit-day07} shows that: (1) users who revisited by day $t$ tend to have more days active in the following 1-month period compared to users who did not revisit by day $t$ (distribution less left skewed); (2) users who revisited sooner tend to have more days active in the following 1-month period. It is worth mentioning that the fact that they revisit soon does not imply that they will be more active. Instead, it could also be true that users who are more active tend to come back sooner. 

\begin{figure}[t]
  \centering
  \includegraphics[width=0.95\linewidth]{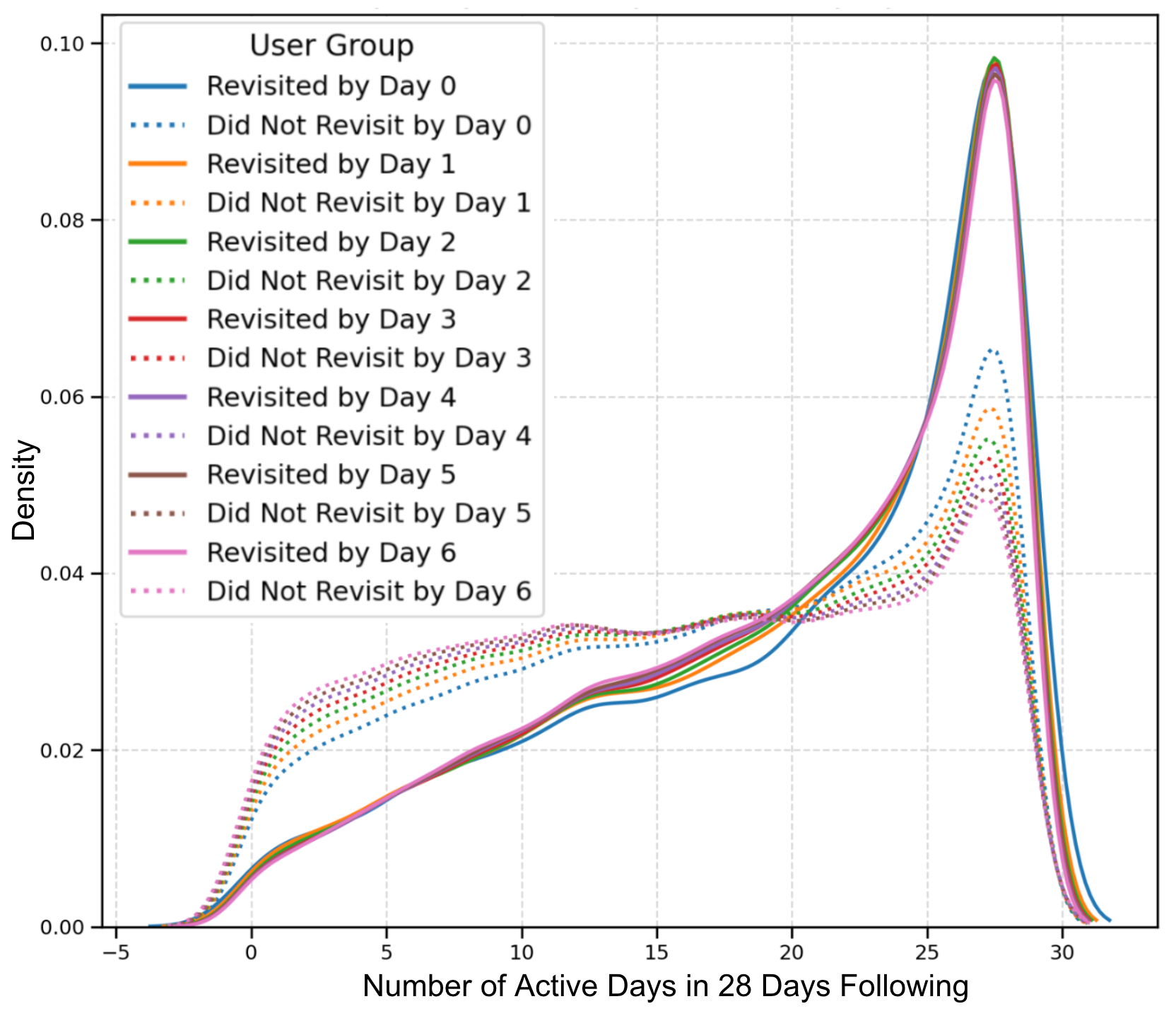}
  \caption{28-day activity distribution by revisit status (revisit by day 0-6)}
\label{28d-revisit-day07}
\end{figure}

\item \textbf{Revisitation grid-click drives deeper engagement than just revisitation impression.} We have one group of users with only revisitation impressions and another group of users with revisitation grid-click. We computed the difference in their active days between the two groups in the 1-month period after the initial revisitation by Day 0 to Day 6 and estimate their incremental change within the 1-month period. 
Figure \ref{corre} shows that \textbf{Grid-click-based revisitation} has a higher correlation with the change in the 1-month activity than pure Impression-based revisitation. Unsurprisingly, this suggests that more ``intentional" revisitation (i.e. via a grid click over just an impression) yields superior retention outcomes. 
\begin{figure}[t]
  \centering
  \includegraphics[width=\linewidth]{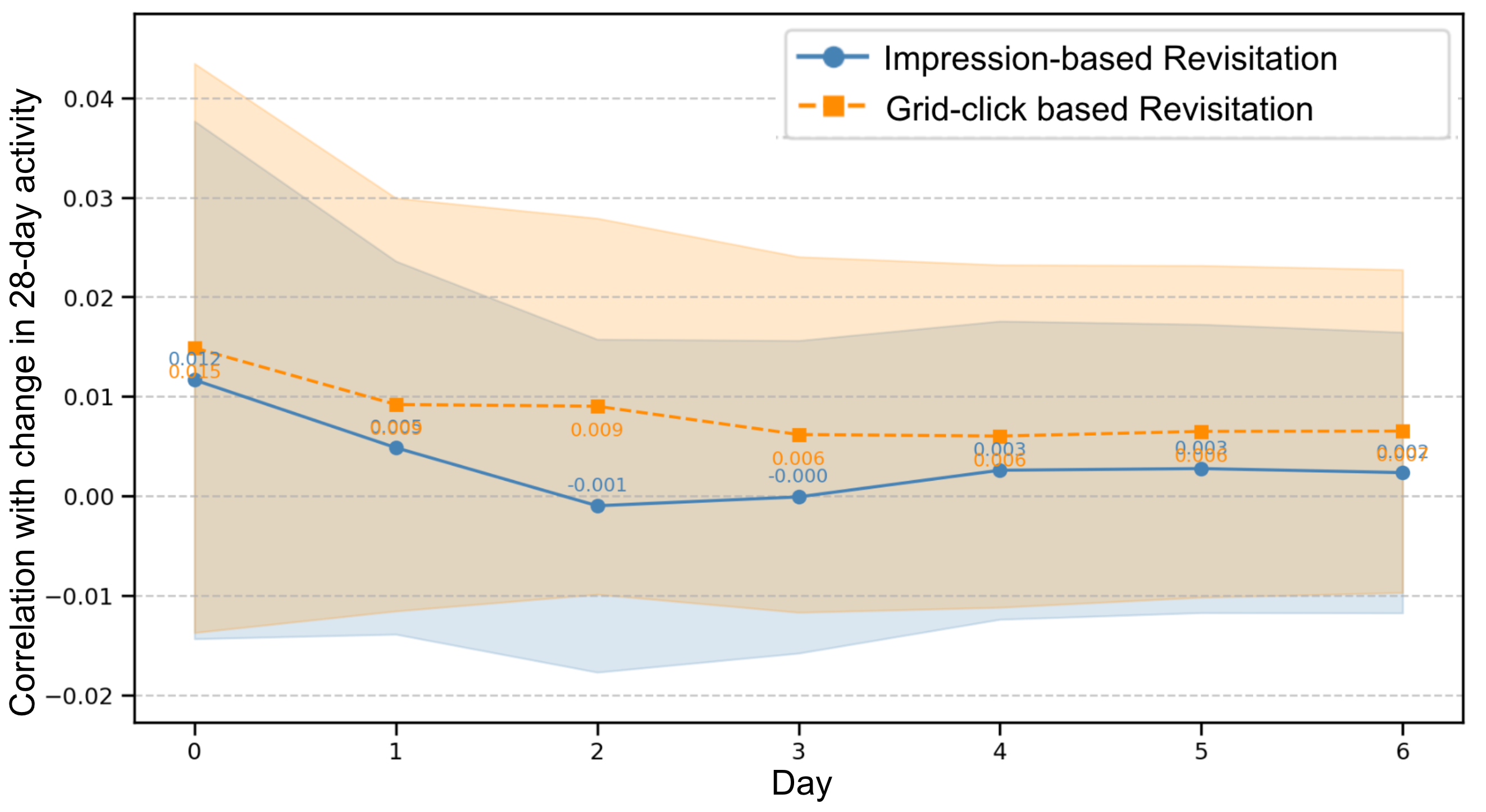}
  \caption{Correlation Between Revisit Type by Day X and $\Delta$ in 28-day Engagement with 95\% Confidence Intervals}
\label{corre}
\end{figure}

\end{enumerate}

\section{Revisitation Modeling in Multi-task Learning Recommender System}

\begin{figure}[t]
     \centering
\includegraphics[width=.75\linewidth]{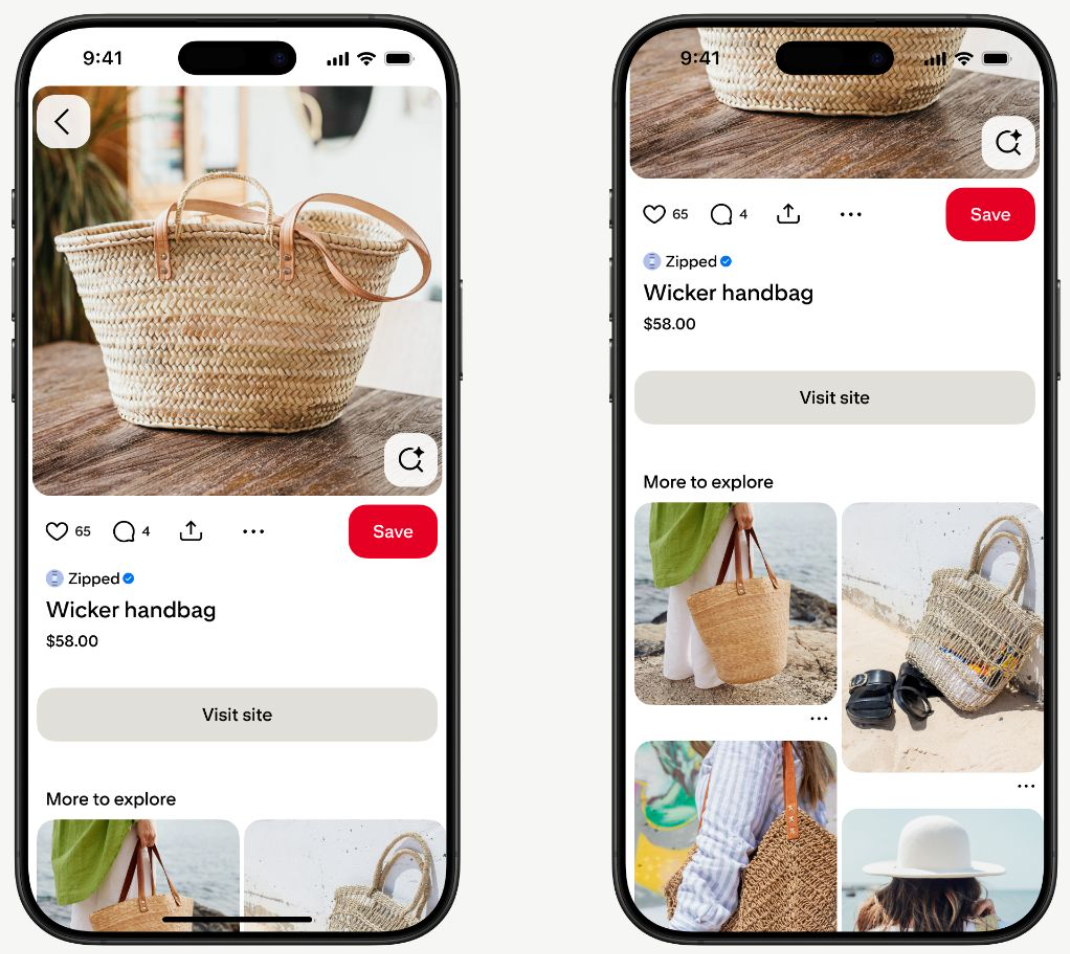}
     \caption{Related Pins Surface on Pinterest}
     \label{rp}
\end{figure}

We proposed revisitation modeling on the Related Pins surface on Pinterest \cite{liu2017related} (Figure \ref{rp}), a search-like surface which serves around 50\% of the traffic on the entire platform. Users enter the Related Pins surface when they grid-click a Pin on any other upper-stream surface (such as Homefeed or Search). There are two main parts of the Related Pins surface: (1) A closer view of the Pin (query Pin) they just grid-clicked on from an upper-stream surface where they can read the title and descriptions of the Pin, and save the Pin to their own profile so they can revisit it any time later (Figure \ref{rp} left). (2) When user scrolls down, they will see a grid view of recommended Pins under title ``More like this" which are related to the query Pin above (Figure \ref{rp} right). Users can engage with the the recommended Pins, a.k.a. candidate Pins with the following main actions:
\begin{itemize}
\item {\textbf{Impression}}: Scroll or navigate through a Pin without deeper engagement.
\item {\textbf{Grid-click}}: A single tap on the Pin that makes it appear larger. When a grid-click happens, user will enter another Related Pin feed where the query Pin becomes the previously tapped Pin and new recommendations will be generated for that Pin.
\item {\textbf{Save (a.k.a. Repin)}}: Click on the ``Save" button to save the Pin to user's own profile so they can revisit it any time later. 
\item {\textbf{Click}}: Click on the ``Visit" button to view more information about the Pin from a third-party website (if there is the Visit botton for that Pin). 
\item {\textbf{Long click}}: User stay on the third-party website for longer than 35s.
\end{itemize}

\subsection{Multi-task Learning Recommender System}

\begin{figure}[t]
  \centering
  \includegraphics[width=\linewidth]{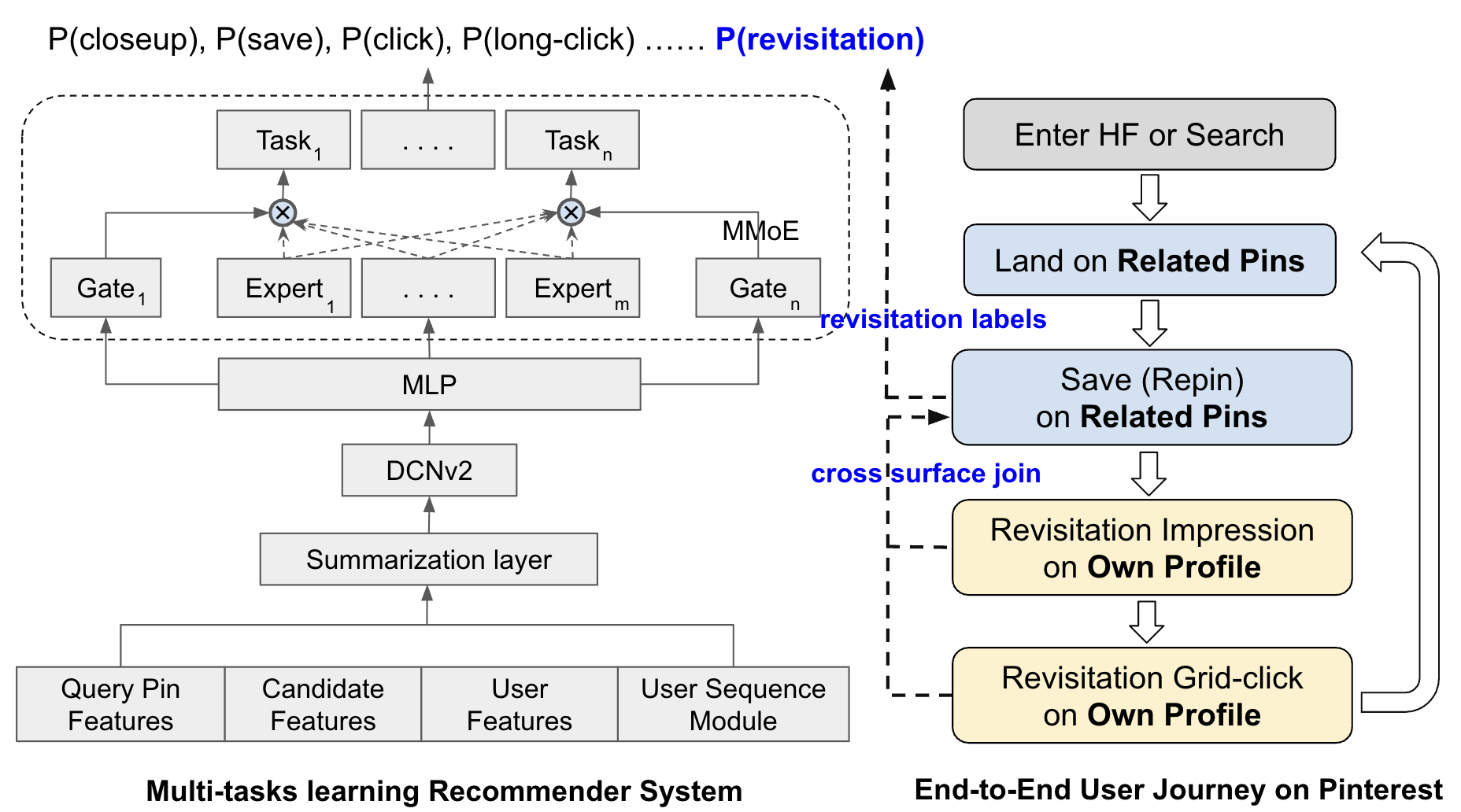}
  \caption{Revisitation Data Generation and Modeling in Multi-task Learning Recommender System}
\label{model}
\end{figure}

 The goal of a multi-task recommender system is to jointly optimize for multiple related tasks by leveraging shared representations and correlations among those tasks \cite{zhang2021survey}.
 Figure \ref{model} (left) shows the main architecture of the multi-task learning ranking model for recommendation on Related Pins. The input layer includes query Pin features, candidate Pin features, user features, and a transformer-based user sequence module that encodes user historical actions. The summarization layer includes some feature crossing manipulation between query Pin features and candidate Pin features and dimension reductions, where the output embedding is fed to a DCNv2 module \cite{wang2021dcn} to learn deeper feature interactions. A final MLP layer aggregates the output embeddings of DCNv2 into a single embedding and feed it to an MMoE module \cite{ma2018modeling} for multi-task learning. 
 Each action is binary so the loss function is the sum of binary cross entropy loss (Equation (\ref{loss})), 
 
\begin{equation}
  \text{loss}_1 = \sum_{i\in\mathbb{L}} w_i \cdot [c_i \text{log} \,\text{p}(c_i|q, \theta) + (1-c_i) \text{log} \,\text{p}(1-c_i|q, \theta)]
  \label{loss}
\end{equation}
where $\mathbb{L}=\{grid\text{-}click, \space repin, \space click, \space long click\}$
 , $q$ refers to query Pin, $c$ refers to candidate Pin, $\theta$ denotes model parameters, $c_i$ is binary label, where 1 indicating action $i$ is positive and 0 means negative, and $w_i$ refers to the loss weight for task $i$.

The recommender system requires an aggregated score based on the prediction of each task to rank all the candidate Pins. For each task, a utility weight $u_i$ representing the importance of the action based on business need is tuned and assigned to $p(c_i|q)$ and all the tasks are aggregated to calculate the final score for ranking the candidate Pins (Equation (\ref{score})). 

 \begin{equation}
  \text{score} = \sum_{i\in\mathbb{L}} u_i  \cdot \text{p}(c_i|q, \theta)
  \label{score}
\end{equation}

\subsection{Revisitation Label Design}
\textit{Goal.} We seek a revisitation signal that balances label density (coverage) and intent (precision) as the best proxy for retention.

\textit{Interaction type.} Impression-based revisits provide high  coverage but include incidental scroll noise; grid‑click revisits are rarer but indicate deliberate return intent. Section ``Revisitation Behavior Analysis" shows grid‑click revisits correlate more strongly with downstream activity than impressions.

\textit{Temporal window.} In section ``Revisitation Behavior Analysis", we observed that revisits concentrate shortly after saves (sharp decay within the first week). In addition, the sooner revisitation happens, the more likely it is for users to be more active within the following 1-month period. Therefore, we utilized the revisitation behaviors that happened 0-6 days after repin as revisitation labels as they are more recent and take up most of the revisitation actions. We surmised that driving more revisitation behaviors on the saved Pin in the following week could lead to an increase in active days by users and potentially improve user retention. In addition, based on our finding that revisitation grid-clicks take less than half of the volume of revisitation impressions but they tend to drive deeper engagement than revisitation impressions, 
we utilized revisitation grid-clicks that happened 0-6 days after repin while only use same‑day revisitation impressions as repin. The final revisitation labels include:
\begin{itemize}
    \item \textbf{Same-day Revisitation Impression (1dRevImpre)}: User saved a Pin and has a revisitation impression of the saved Pin on the same day. The volume of same-day revisitation impressions consists of 25.8\% of the repin volume in a day.

    \item \textbf{Same-day Revisitation Grid-click (1dRevGrid)}: User saved a Pin and has a revisitation grid-click of the saved Pin on the same day. The volume of same-day revisitation grid-clicks takes up 4.7\% of the repin volume in a day.
    
    \item \textbf{7-day Revisitation Grid-click (7dRevGrid)}: User saved a Pin and has a revisitation grid-click of the saved Pin in the following 0-6 days. The volume of 7-day revisitation grid-click consists of 9\% of the repin volume in a day in total as we see a quick decay of revisitation grid-click after repin on day 0 in Figure \ref{b}.

\end{itemize}

We merge the three types of labels as the final revisitation label, and add a revisitation head in the multi-tasks learning ranking model for predicting the probability of a candidate Pin being saved and revisited in the future using the constructed revisitation label. Therefore, the loss function with the additional revisitation head is shown in Equation (\ref{loss1}), where $RP$ indicates $Repin$ and $RV$ stands for $Revisit$.
\begin{multline}
  \text{loss}_2 = \text{loss}_1 + 
  w_{RP\&RV} \cdot [c_{RP\&RV}\text{log} \,\text{p}(c_{RP\&RV}|q, \theta)\\ + (1-c_{RP\&RV}) \text{log} \,\text{p}(1-c_{RP\&RV}|q, \theta)]
  \label{loss1}
\end{multline}

The final ranking score is shown in Equation (\ref{score1}), and we tune the utility weight $u_{RP\&RV}$ for the new task.

 \begin{equation}
  \text{score} = \sum_{i\in\mathbb{L}} u_i  \cdot \text{p}(c_i|q, \theta) + u_{RP\&RV}  \cdot \text{p}(c_{RP\&RV}|q, \theta)
  \label{score1}
\end{equation}

\subsection{Revisitation Features Design}

The current recommender system on the Related Pins surface takes various Pin features as input, including content features, such as text embeddings and visual embeddings \cite{zhai2019learning}, and graph based embeddings, such as \textit{graphsage} embeddings \cite{hamilton2017inductive}, \textit{pinnersage} embeddings \cite{pancha2022pinnerformer}, \textit{omnisage} embeddings \cite{badrinath2025omnisage}, and engagement features like \textit{Navboost} \cite{kislyuk2015human}. One of the most important engagement features that have been adopted on all the surfaces is called Pin perf, counting the historical number of actions on a specific Pin. For each positive action (including grid-click, save, click, and long click), the Pin perf feature counts the how many times a Pin has been engaged with that action on Pinterest in the past 90 days. 

In this work, we designed a set of revisitation Pin perf features to evaluate the popularity of the saved Pins receiving revisitations in the past 7 days, 30 days, and 90 days. Corresponding to the three types of revisitation labels, we designed three types of revisitation Pin perf features, which we refer as Related Pins-triggered revisitation Pin perf features since we only count +1 if the revisited Pin is saved on the Related Pins surface by the same user. This design is for enhancing the model learning correlations between a Pin's revisitation popularity in history (features) and its current revisitation performance (labels). We aggregated the features on the 7-day (update very day), 30-day (update every 3 days), and 90-day (update every week) basis for the sake of higher feature coverage. 
\begin{itemize}
\item Same-day Revisit Impression Pin Perf: $\sim 60\%$ coverage
\item Same-day Revisit Grid-click Pin Perf: $\sim 49\%$ coverage
\item 7-day Revisit Grid-click Pin Perf: $\sim 51\%$ coverage

\end{itemize}

    


In addition, to capture the general popularity of a Pin being revisited, we designed a set of overall revisitation Pin perf features where we do not require the repin happened on the Related Pin surface and do not define the time gap between repin and revistation.
\begin{itemize}
\item Overall Revisit Impression Pin Perf: $\sim 80\%$ coverage
\item Overall Revisit Grid-click Pin Perf: $\sim 70\%$ coverage
\end{itemize}

For each Pin perf feature, we include a count of ``number of actions" and a count of ``unique users" to reflect volume and popularity, respectively. Volume alone could be noise to the system since the Pin revisitation count could be heavily driven by a single enthusiastic user.

\subsection{Cross-surface Revisitation Data Pipeline}

 Figure \ref{model} (right) illustrates the end-to-end user journey on Pinterest. When user lands on Related Pins surface from other surface (such as Homefeed and Search) and grid-clicks a Pin they like, they can save (repin) the Pin to their own profile. Later on, they can go back to their own profile and browse the Pin (revisitation impression); they can also tap on the saved Pin to have a bigger and deeper view of the Pin (revisitation grid-click) where they will re-land on Related Pins with the saved Pin as query Pin and browse the recommended Pins that are to it. Users' actions on all the candidate Pins in the grid view on the Related Pins surface (staging 2 and 3 with blue color) are logged to a daily generated dataset for training the ranking model for recommendations on the Related Pins surface. However, users' revisitation actions on their own profiles were not included in the training data. In order to further track users' revisitation behaviors on their saved Pins, we did a cross-surface join of the existing training data of user actions on the Related Pins surface and data of user actions on their own profile on user ID and Pin ID to map users' revisitation actions in their own profile to their saved Pins on related Pins surface. 

 \begin{figure}[t]
  \centering
  \includegraphics[width=\linewidth]{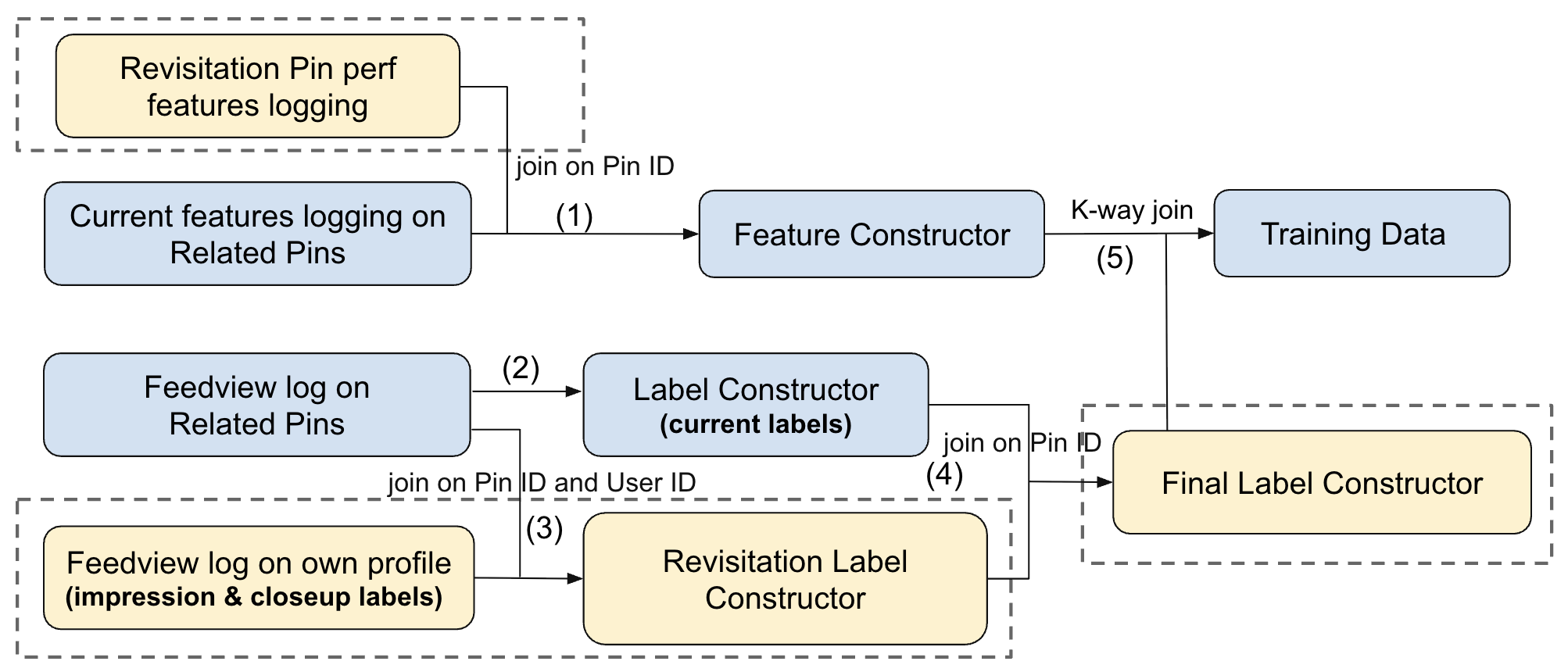}
  \caption{Revisitation Data Generation and Training Data Generation Pipeline}
\label{data}
\end{figure}

Figure \ref{data} illustrates the revisitation data generation and the final training data generation pipeline. 
\begin{enumerate}
    \item  Log the revisitation Pin perf features, join them with the other features, and convert raw features log entries into TabularML.
    \item  Extracts all the user actions on the Related Pins surface from Feedview Log into labels.
    \item Cross-surface and cross-session (day) join of users' save actions on the Related Pins surface and users' revisitation impressions and revisitation grid-clicks actions on their own profile (join key: User ID and Pin ID, condition: $time_{save} < time_{revisit} < time_{save+7(days)}$). 
    \item  Join the revisitation labels with the repin labels and on User ID and Pin ID and add a new column for the revisitation label  indicating which saved Pin has been revisited in the final label constructor.
    \item  Join feature TabularML and the final labels (with revistation label added) on API request id and candidate Pin id. 
\end{enumerate}

Step (1) (2) and (3) can be conducted in parallel, followed by step (4) and step (5) sequentially. This solution of mapping user revisitation labels to training data is scalable to any other surface on Pinterest and other platforms where users can save content and revisit the saved content.

\section{Offline Experiment}

We trained the multi-task learning ranking model on the Related Pins surface on Pinterest illustrated in Figure \ref{model} with the designed revisitation labels and revisitation Pin perf features. We used 27 days of data for training, the last day of training data for calibration and the following 3 days of data for evaluation. 
The size of the entire training data is around 6.6 billion and the size of the evaluation data is around 700 million. We train the model for 1 epoch. We also tuned the utility weight $u_{RP\&RV}$ for the added revisitation task in Equation (\ref{score1}) and experiment results showed that the model achieved the best gain without tradeoff when $u_{RP\&RV} = 1.27*u_{Repin}$.

Table \ref{offline} first shows the offline evaluation results. For ranking metrics per task (head), we observed revisitation head achieved 20\%-40\% lift and repin head also achieved 0.05\%-0.03\% lift compared to the current ranking model without revisitation modeling. Because all the revisited Pins were saved Pins (repins), the model was able to prioritize ranking the Pins that may receive more revisitations in the top positions. 

In addition, we used \textit{Hits@3} to evaluate the overall ranking performance based on the final ranking score defined in Equation (\ref{score1}). For example, we incremented \textit{Hits@3 (Repin)} if the user saved any of the top 3 recommendations. We tuned the utility weight for the revisitation head in Equation (\ref{score1}) based on \textit{Hits@3} metrics. By modeling revisitation in the ranking model, the model improved the \textit{Hits@3} metrics on repin by $0.59\%$ and revisitation by $0.65\%$ significantly without trade-offs on other heads. 
Empirical evidence at Pinterest showed that \textit{Hits@3} usually lines up the most with the online A/B experiment among all of the metrics. 

\begin{table}[t]
\begin{threeparttable}
  \caption{Offline Evaluation Results}
  \label{offline}
  \begin{tabular}{ccccccccc}
    \toprule
    Metric & Repin&Revisit\\
    \midrule
    NDCG@3 Lift (\%) &  \textbf{0.1279} & \textbf{40.15}  \\             
    
    MAP@3 Lift (\%)& \textbf{0.06453}& \textbf{43.44}\\
    
    Recip\_Rank@3 Lift (\%)	&-0.02666	& \textbf{27.62} \\
    
    Recall@3 Lift (\%)& \textbf{0.2705}	& \textbf{35.4} \\
    
    Pariwise Accuracy Lift (\%)& \textbf{0.10} &	\textbf{21.62} \\
    \midrule
    Hits@3 Lift (\%) & 	\textbf{0.59} & \textbf{0.65}  \\
  \bottomrule
\end{tabular}
\begin{tablenotes}
      \small
      \item Note: Bold numbers indicate metrics gain.
    \end{tablenotes}
    
\end{threeparttable}
\end{table}

\section{Online A/B Experiment}

We conducted a large-scale online A/B experiment from April 29, 2025 to June 26, 2025 on Pinterest Related Pins surface \cite{liu2017related}. The number of users in each group was around 12 million. The experiment ran for 2 months from which we reported the accumulated results. 



\subsection{Revisitation Online Metrics}

Table \ref{online3} first shows the three types of revisitation metrics corresponding to the proposed revisitation labels, i.e., 1dRevImpre, 1dRevGrid, and 7dRevGrid. 
Compared to the control group, the treatment group achieved 0.95\% - 1.42\% lift on the revisitation grid-clicks metrics. The revisitation grid-clicks metrics gains are higher than the repin gain, which indicating revisitation grid-click rate gain overall after removing the impact of repin gain on revisitation metrics.
Notably, the treatment group achieved higher revisitation ratio in terms of both volume and propensity, indicating that the revisitation modeling in the treatment group was driving larger proportion of revisitations from the saved Pins. Compared to the same-day revisitation grid-click metrics, the treatment group achieved higher lift on the 7-day revisitation grid-clicks metrics, which demonstrated strong accumulative effect of revisitation modeling in driving users' longer term revisitation behaviors on the platform.

We also observed that the gain on the same-day revisitation impression is lower than the grid-click based revisitation metrics. 
This aligns with our previous finding that revisitation grid-clicks drives deeper engagement than just Impression-based revisitation. 

\subsection{Engagement Metrics and User Retention Metrics} 
Table \ref{online3} also shows the engagement metrics lift on the Related Pins surface between the control group and the treatment group with revisitation modeling. Aligned with the offline metrics, the treatment group achieved better metrics of $0.94\%$ volume gain and  $0.64\%$ propensity gain on repin.
Table \ref{online3} showed that the treatment group with revisitation modeling is driving $+0.41\%$ user sessions that are longer than 5 minutes and $+0.35\%$ Web or API requests overall on Pinterest. 


We also observed from Table \ref{online3} that the users in the treatment group spent longer time on the Related Pins surface ($+0.53\%$) as well as their own profile ($+0.68\%$) where revisitations happen, and overall in App ($+0.39\%$) than users in the control group. Given that our model only directly affected the recommendations on the Related Pins surface, the gains on site-wide retention metrics were considered as significantly notable. All these metrics gains contributed to $0.1\%$ volume gain and $0.08\%$ propensity gain on active users.

\begin{table}[t]
  \caption{Online Experiment Results Lift (\%)}
  \label{online3}
  \begin{tabular}{ccc}
    \toprule
    Metric Name &\makecell{Actions Per User \\(Volume)}& \makecell{Unique Users \\ (Propensity)}\\
    \midrule
    1dRevImpre & 0.65** & 0.93**\\
    1dRevGrid & 0.95** & 1.17**\\
7dRevGrid & 1.18** & 1.42** \\  
\midrule
    Repin & 0.94* & 0.64***\\
    Sessions $\geq$ 5min & 0.41 *** & -0.03\\
    Web Or API Requests & 0.35 *** & 0.00 \\
    \hline
    Time Spent (RP) & 0.53 *** & 0.02\\
    Time Spent (OP) & 0.68** & 0.13\\
    \makecell{Total Time Spent} & 0.39 *** & 0.01\\
    Active Users & 0.10** & 0.08**\\
  \bottomrule
\end{tabular}
\begin{tablenotes}
    \item RP: Related Pins surface, OP: own profile
    \item ***: p$<$0.01, **: p$<$0.05, *: p$<$0.1
\end{tablenotes}
\end{table}


\subsection{Visualization}
Based on the online A/B experiment results, we further conducted multiple revisitation visualizations on the 16 main topics of Pins on Pinterest.

\subsubsection{Interpretability}
\begin{figure}[t]
  \centering
  \includegraphics[width=\linewidth]{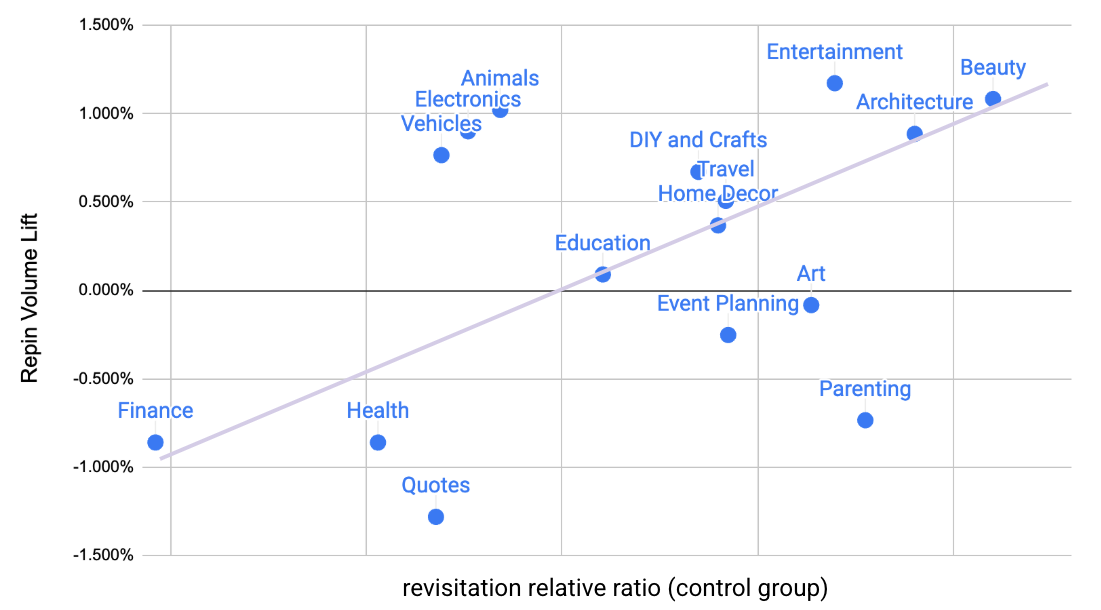}
  \caption{Interpretability (x-axis: revisitation relative ratio of control group (absolute value hidden while order preserved),  y-axis: repin volume lift \%)}
\label{reasoning}
\end{figure}

By incoporating revisitation modeling into the multi-task recommendation system (Equation (\ref{loss1})), the model is prioritizing recommending Pins that achieved more saves and revisitations. Pins that received more revisitations historically will tend to receive more repins by the proposed revisitation modeling framework. We calculated the average probability of repin and revisitation $P(c_{RP \& RV}|q, \theta)$ on each topic for the control group, and the repin volume lift (\%) per user of treatment group over control group. Figure \ref{reasoning} shows that the topics that have the highest $P(c_{RP \& RV}|q, \theta)$ (such as \textit{Beauty}, 
\textit{Architecture}, and \textit{Entertainment}) generated higher repin volume lift than other topics in the treatment group, while the topics that have the lowest $P(c_{RP \& RV}|q, \theta)$ (e.g., \textit{Finance}, \textit{Health}, and \textit{Quotes}) incurred more negative repin lift. 

\subsubsection{Repin Rate v.s. Revisitation Rate}
\begin{figure}[t]
  \centering
  \includegraphics[width=0.9\linewidth]{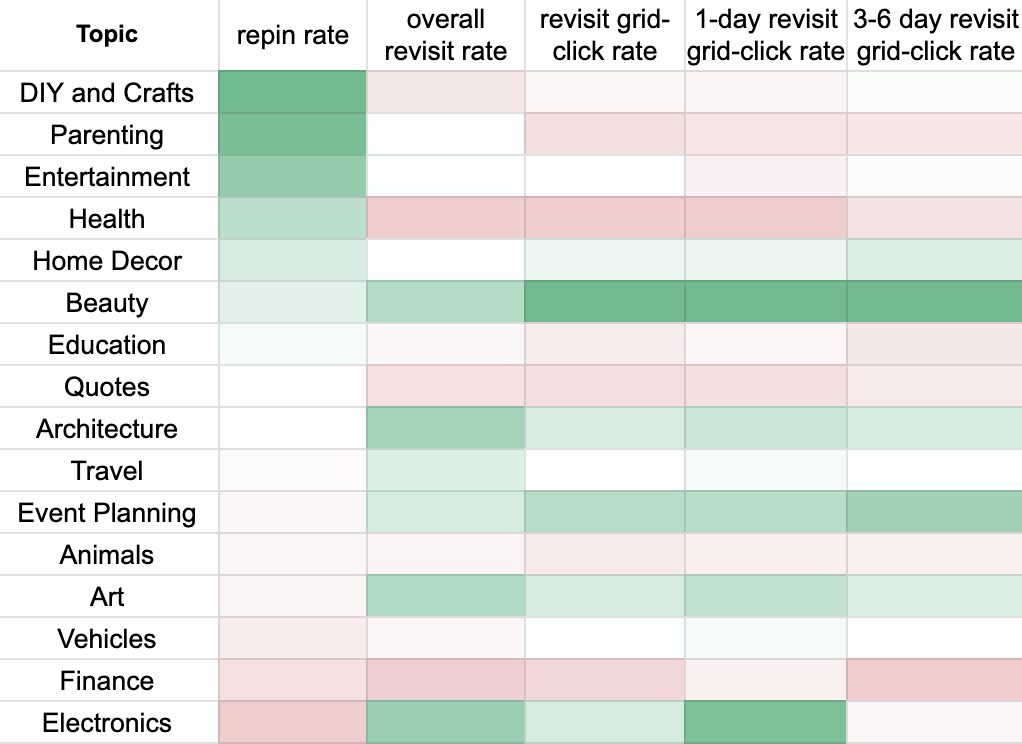}
  \caption{Heatmap of Repin Rate v.s. Revisitation Rate by Topic (Green represents higher rate and red vice versa)}
\label{repin_revisit}
\end{figure}

Figure \ref{repin_revisit} uses a heatmap to illustrate the average repin rate, average overall revisitation rate, and revisitation grid-click rates for each topic. We observed that topics such as \textit{DIY and Crafts}, \textit{Parenting}, and \textit{Health} which requires more real-life practices received higher repin rates but lower revisitation rates. On the other hands, \textit{Beauty}, \textit{Architecture}, \textit{Travel}, \textit{Event Planning}, \textit{Art}, \textit{Electronics} received lower repin rates but high revisitation rates. 

\subsubsection{Long-term  v.s. Short-term Revisitation}
In Figure \ref{b}, we observed that the volume of revisitation grid-clicks decays dramatically in the first three days. 8\% of the saved Pins received a revisitation grid-click after 0-2 days while the ratio dropped to only 1.9\% in the next four days (day 3 to day 6). In Table \ref{v1}, we calculated $revisit_{3-6 day}$, the ratio of 3-6 day revisitation grid-click volume to the 0-6 day revisitation grid-click volume for each topic, where a higher ratio indicates that the topic tends to be revisited 3-6 days after saving (longer term revisitation interests). The longer term revisitation interests are \textit{Event Planning}, \textit{Health}, \textit{Home Decor}, and \textit{DIY and Crafts}, which are innately related to long-term user interests. The topics which users have the most short-term revisitation interest are \textit{Finance}, \textit{Electronics}, \textit{Vehicles}, \textit{Architecture}, and \textit{Art}. 

\begin{table}[ht]
  \caption{Long-term  v.s. Short-term Revisitation by Topic}
  \label{v1}
  \begin{tabular}{cc|cc}
  \toprule
  Topic &  ratio & Topic &  ratio \\
  \midrule
  Event Planning & 0.1881 &  Entertainment & 0.1639\\
  Health & 0.1836 & Animals & 0.1633 \\ 
  Home Decor & 0.1780 & Education & 0.1602 \\
  DIY and Crafts & 0.1733 & Art & 0.1572 \\
  Quotes & 0.1725 &  Architecture & 0.1495\\
  Beauty & 0.1666 & Vehicles & 0.1334\\
  Parenting & 0.1659 & Electronics & 0.1079\\
  Travel & 0.1654 & Finance & 0.1056\\
  \bottomrule
  \end{tabular}
\end{table}


 
\section{Conclusion}
This work introduces a revisitation modeling framework in a multi-task learning recommender system, using Pinterest as a test bed. By analyzing user save (repin) actions and their cross-session, cross-date revisitation behaviors, we establish causal links between saved Pins and user retention. Our approach encompasses comprehensive behavior analysis, label and feature design for revisitation, and demonstrates scalability to other online platforms. Differing from traditional recommender systems that rely on in-session actions, our model leverages cross-session, cross-date, and cross-surface revisitation data, aiming to enhance long-term user retention and core engagement metrics, as validated through both offline and large-scale online experiments.

While our results show the framework effectively drives engagement, revisitation, and retention metrics, we identified a bottleneck: highly active users contribute disproportionately by repinning more than less active users, leading to limited revisitation label data from the latter group. To address this, future work could broaden revisitation actions beyond explicit repin events to include implicit signals, such as users interacting with Pins on similar topics across consecutive sessions. Overall, our work offers a novel, scalable framework for estimating and optimizing user revisitation and long-term retention.

\section{Acknowledgments}
We would like to acknowledge our cross‑team colleagues Jing Zhang and Yinuo Liu for their invaluable contributions to the productionization of the revisitation labels and revisitation features, respectively. We are grateful to Haoran Guo and Liyao Lu for their excellent guidance in shipping this work to production. We also thank Shivin Thukral, Tianyu Feng, Wendy Shao, Jinfeng Rao, Lu Liu, Jinyu Xie, Matt Chun, Michael Chau, Karim Wahba, and Rajat Raina for their generous support throughout this work.

\bigskip

\bibliography{aaai2026}

@inproceedings{wang2022surrogate,
  title={Surrogate for long-term user experience in recommender systems},
  author={Wang, Yuyan and Sharma, Mohit and Xu, Can and Badam, Sriraj and Sun, Qian and Richardson, Lee and Chung, Lisa and Chi, Ed H and Chen, Minmin},
  booktitle={Proceedings of the 28th ACM SIGKDD conference on knowledge discovery and data mining},
  pages={4100--4109},
  year={2022}
}

@inproceedings{wang2021dcn,
  title={Dcn v2: Improved deep \& cross network and practical lessons for web-scale learning to rank systems},
  author={Wang, Ruoxi and Shivanna, Rakesh and Cheng, Derek and Jain, Sagar and Lin, Dong and Hong, Lichan and Chi, Ed},
  booktitle={Proceedings of the web conference 2021},
  pages={1785--1797},
  year={2021}
}

@inproceedings{ma2018modeling,
  title={Modeling task relationships in multi-task learning with multi-gate mixture-of-experts},
  author={Ma, Jiaqi and Zhao, Zhe and Yi, Xinyang and Chen, Jilin and Hong, Lichan and Chi, Ed H},
  booktitle={Proceedings of the 24th ACM SIGKDD international conference on knowledge discovery \& data mining},
  pages={1930--1939},
  year={2018}
}

@article{zhang2021survey,
  title={A survey on multi-task learning},
  author={Zhang, Yu and Yang, Qiang},
  journal={IEEE transactions on knowledge and data engineering},
  volume={34},
  number={12},
  pages={5586--5609},
  year={2021},
  publisher={IEEE}
}

@inproceedings{zhao2023user,
  title={User retention-oriented recommendation with decision transformer},
  author={Zhao, Kesen and Zou, Lixin and Zhao, Xiangyu and Wang, Maolin and Yin, Dawei},
  booktitle={Proceedings of the ACM Web Conference 2023},
  pages={1141--1149},
  year={2023}
}

@inproceedings{ding2023interpretable,
  title={Interpretable user retention modeling in recommendation},
  author={Ding, Rui and Xie, Ruobing and Hao, Xiaobo and Yang, Xiaochun and Ge, Kaikai and Zhang, Xu and Zhou, Jie and Lin, Leyu},
  booktitle={Proceedings of the 17th ACM Conference on Recommender Systems},
  pages={702--708},
  year={2023}
}

@inproceedings{cai2023reinforcing,
  title={Reinforcing user retention in a billion scale short video recommender system},
  author={Cai, Qingpeng and Liu, Shuchang and Wang, Xueliang and Zuo, Tianyou and Xie, Wentao and Yang, Bin and Zheng, Dong and Jiang, Peng and Gai, Kun},
  booktitle={Companion Proceedings of the ACM Web Conference 2023},
  pages={421--426},
  year={2023}
}

@article{afsar2022reinforcement,
  title={Reinforcement learning based recommender systems: A survey},
  author={Afsar, M Mehdi and Crump, Trafford and Far, Behrouz},
  journal={ACM Computing Surveys},
  volume={55},
  number={7},
  pages={1--38},
  year={2022},
  publisher={ACM New York, NY}
}

@inproceedings{jia2025learn,
  title={LEARN: Knowledge Adaptation from Large Language Model to Recommendation for Practical Industrial Application},
  author={Jia, Jian and Wang, Yipei and Li, Yan and Chen, Honggang and Bai, Xuehan and Liu, Zhaocheng and Liang, Jian and Chen, Quan and Li, Han and Jiang, Peng and others},
  booktitle={Proceedings of the AAAI Conference on Artificial Intelligence},
  volume={39},
  number={11},
  pages={11861--11869},
  year={2025}
}

@inproceedings{gong2025multiple,
  title={Multiple Purchase Chains with Negative Transfer Elimination for Multi-Behavior Recommendation},
  author={Gong, Shuwei and Liu, Yuting and Dang, Yizhou and Guo, Guibing and Zhao, Jianzhe and Wang, Xingwei},
  booktitle={Proceedings of the AAAI Conference on Artificial Intelligence},
  volume={39},
  number={11},
  pages={11717--11725},
  year={2025}
}

@inproceedings{dang2025augmenting,
  title={Augmenting Sequential Recommendation with Balanced Relevance and Diversity},
  author={Dang, Yizhou and Zhang, Jiahui and Liu, Yuting and Yang, Enneng and Liang, Yuliang and Guo, Guibing and Zhao, Jianzhe and Wang, Xingwei},
  booktitle={Proceedings of the AAAI Conference on Artificial Intelligence},
  volume={39},
  number={11},
  pages={11563--11571},
  year={2025}
}

@inproceedings{guo2022intelligent,
  title={Intelligent online selling point extraction for e-commerce recommendation},
  author={Guo, Xiaojie and Wang, Shugen and Zhao, Hanqing and Diao, Shiliang and Chen, Jiajia and Ding, Zhuoye and He, Zhen and Lu, Jianchao and Xiao, Yun and Long, Bo and others},
  booktitle={Proceedings of the AAAI Conference on Artificial Intelligence},
  volume={36},
  number={11},
  pages={12360--12368},
  year={2022}
}

@inproceedings{wang2020discovery,
  title={Discovery news: a generic framework for financial news recommendation},
  author={Wang, Chong and Kim, Lisa and Bang, Grace and Singh, Himani and Kociuba, Russell and Pomerville, Steven and Liu, Xiaomo},
  booktitle={Proceedings of the AAAI Conference on Artificial Intelligence},
  volume={34},
  number={08},
  pages={13390--13395},
  year={2020}
}

@inproceedings{gugnani2020implicit,
  title={Implicit skills extraction using document embedding and its use in job recommendation},
  author={Gugnani, Akshay and Misra, Hemant},
  booktitle={Proceedings of the AAAI conference on artificial intelligence},
  volume={34},
  number={08},
  pages={13286--13293},
  year={2020}
}

@article{badrinath2025omnisage,
  title={OmniSage: Large Scale, Multi-Entity Heterogeneous Graph Representation Learning},
  author={Badrinath, Anirudhan and Yang, Alex and Rajesh, Kousik and Agarwal, Prabhat and Yang, Jaewon and Chen, Haoyu and Xu, Jiajing and Rosenberg, Charles},
  journal={arXiv preprint arXiv:2504.17811},
  year={2025}
}

@inproceedings{pancha2022pinnerformer,
  title={Pinnerformer: Sequence modeling for user representation at pinterest},
  author={Pancha, Nikil and Zhai, Andrew and Leskovec, Jure and Rosenberg, Charles},
  booktitle={Proceedings of the 28th ACM SIGKDD conference on knowledge discovery and data mining},
  pages={3702--3712},
  year={2022}
}

@article{hamilton2017inductive,
  title={Inductive representation learning on large graphs},
  author={Hamilton, Will and Ying, Zhitao and Leskovec, Jure},
  journal={Advances in neural information processing systems},
  volume={30},
  year={2017}
}

@inproceedings{zhai2019learning,
  title={Learning a unified embedding for visual search at pinterest},
  author={Zhai, Andrew and Wu, Hao-Yu and Tzeng, Eric and Park, Dong Huk and Rosenberg, Charles},
  booktitle={Proceedings of the 25th ACM SIGKDD international conference on knowledge discovery \& data mining},
  pages={2412--2420},
  year={2019}
}

@inproceedings{liu2017related,
  title={Related pins at pinterest: The evolution of a real-world recommender system},
  author={Liu, David C and Rogers, Stephanie and Shiau, Raymond and Kislyuk, Dmitry and Ma, Kevin C and Zhong, Zhigang and Liu, Jenny and Jing, Yushi},
  booktitle={Proceedings of the 26th international conference on world wide web companion},
  pages={583--592},
  year={2017}
}

@article{kislyuk2015human,
  title={Human curation and convnets: Powering item-to-item recommendations on pinterest},
  author={Kislyuk, Dmitry and Liu, Yuchen and Liu, David and Tzeng, Eric and Jing, Yushi},
  journal={arXiv preprint arXiv:1511.04003},
  year={2015}
}

@inproceedings{zhang2021user,
  title={User Retention: A Causal Approach with Triple Task Modeling.},
  author={Zhang, Yang and Wang, Dong and Li, Qiang and Shen, Yue and Liu, Ziqi and Zeng, Xiaodong and Zhang, Zhiqiang and Gu, Jinjie and Wong, Derek F},
  booktitle={IJCAI},
  pages={3399--3405},
  year={2021}
}

@inproceedings{liu2024modeling,
  title={Modeling User Retention through Generative Flow Networks},
  author={Liu, Ziru and Liu, Shuchang and Yang, Bin and Xue, Zhenghai and Cai, Qingpeng and Zhao, Xiangyu and Zhang, Zijian and Hu, Lantao and Li, Han and Jiang, Peng},
  booktitle={Proceedings of the 30th ACM SIGKDD Conference on Knowledge Discovery and Data Mining},
  pages={5497--5508},
  year={2024}
}

@inproceedings{kwon2020art,
  title={ART (Attractive Recommendation Tailor) How the Diversity of Product Recommendations Affects Customer Purchase Preference in Fashion Industry?},
  author={Kwon, Hyokmin and Han, Jaeho and Han, Kyungsik},
  booktitle={Proceedings of the 29th ACM international conference on information \& knowledge management},
  pages={2573--2580},
  year={2020}
}

@inproceedings{sun2025they,
  title={Why They Come And Go: A Case Study of Productive Flyby Users and Their Rating Integrity Challenge in Movie Recommenders},
  author={Sun, Ruixuan and Kong, Ruoyan and Milton, Ashlee and Kluver, Daniel and Paterson, Ian and Konstan, Joseph A},
  booktitle={Proceedings of the 2025 ACM SIGIR Conference on Human Information Interaction and Retrieval},
  pages={1--11},
  year={2025}
}

@article{jin2017personal,
  title={Personal web revisitation by context and content keywords with relevance feedback},
  author={Jin, Li and Feng, Ling and Liu, Gangli and Wang, Chaokun},
  journal={IEEE Transactions on Knowledge and Data Engineering},
  volume={29},
  number={7},
  pages={1508--1521},
  year={2017},
  publisher={IEEE}
}

@inproceedings{adar2008large,
  title={Large scale analysis of web revisitation patterns},
  author={Adar, Eytan and Teevan, Jaime and Dumais, Susan T},
  booktitle={Proceedings of the SIGCHI conference on Human Factors in Computing Systems},
  pages={1197--1206},
  year={2008}
}

@inproceedings{adar2009resonance,
  title={Resonance on the web: web dynamics and revisitation patterns},
  author={Adar, Eytan and Teevan, Jaime and Dumais, Susan T},
  booktitle={Proceedings of the SIGCHI conference on human factors in computing systems},
  pages={1381--1390},
  year={2009}
}

\end{document}